# Surface texturing of Ti6Al4V alloy using femtosecond laser for superior antibacterial performance

Shazia Shaikh[1], Sunita Kedia[1,*], Deepti Singh[2,3], Mahesh Subramanian[2,3], Sucharita Sinha[1]

[1] Laser & Plasma Surface Processing Section, Bhabha Atomic Research Centre, Mumbai 400085 India.

[2] Bio-Organic Division, Bhabha Atomic Research Centre, Mumbai 400085 India.

[3] Homi Bhabha National Institute, Training School complex, Anushaktinagar, Mumbai 400094 India.

[*] Corresponding authors, Email: skedia@barc.gov.in

**Abstract**

Titanium and its alloy are most widely used implant materials in dental and orthopaedic fields. However, infections occurring during implantation leads to implant failure in most of the cases. Here, we have demonstrated antibacterial behavior of Ti6Al4V alloy achieved when surface modified using femtosecond laser beam. Post laser treatment conical microstructures were observed on the Ti6Al4V alloy surface. Generation of different sub-oxide phases of titanium dioxide were detected on laser treated samples using X-ray diffraction and X-ray photoelectron spectroscopy. Wettability of Ti6Al4V alloy surface changed significantly after interaction with the laser. Adhesion and growth of two gram positive; Staphylococcus aureus and Streptococcus mutans and one gram negative Pseudomonas aeruginosa bacteria have been explored on pristine, as well as, on laser textured Ti6Al4V alloy surfaces. In-vitro investigation on agar plate showed inhibition of bacterial growth on most of the laser treated surface. Superior surface roughness and occurrence of magneli phases of titanium dioxide on laser treated surface were probably responsible for the antibacterial behavior exhibited by the laser treated samples. Therefore, femtosecond laser surface treatment of Ti6Al4V alloy could find potential application in the development of infection free medical implants for dental and orthopedic usages.

**Keyword:** *Ti6Al4V bio-alloy, Femtosecond laser texturing, Antibacterial property.*

## 1. Introduction

Titanium and its alloys have drawn much attention as biomaterials on account of their excellent bio-compatibility and high corrosion resistance in comparison to other metallic biomaterials. Commercially pure titanium and its alloy titanium-aluminum-vanadium (Ti6Al4V) have both been in use for medical applications since 1960 [1]. However, since last few decades Ti6Al4V alloy has largely replaced use of pure titanium due to its superior mechanical strength. This Ti6Al4V alloy has been extensively used as plates, nails, screws and endoprostheses in dental and orthopedic fields [2]. Although, titanium alloy based implants have found wide medical applications, a major drawback has been implant failure caused by contamination and bio-film formation on implant surfaces during implantation necessitating repeated surgeries.

Prevention and proliferation of bacteria on titanium implant surfaces has been a long standing effort in both health care and industrial applications. Basic surface properties such as surface chemistry, wettability, surface topography and roughness have all been reported to influence the attachment of bacteria on implant surfaces [3]. A number of strategies have been adopted to limit bacterial attachment to implant surfaces. The oldest approach was use of biocides which induced death of bacteria. However, extensive use of biocides results in development of higher bacterial resistance to clinically important antibiotics [4]. Another technique has been use of antibacterial metals such as silver, copper, and molybdenum [5-7]. Silver is the oldest and most popular antibacterial metal element. The antibacterial effect of silver arises on its dissolution from bio-implant surface. However, once silver dissolves completely antibacterial property of such an implant diminishes. Use of photo catalytic materials such as titanium dioxide ($TiO_2$) is another approach whereby highly reactive species are formed killing bacteria in implant environment [8]. However, $TiO_2$ having a large band gap requires ultraviolet radiation for effective activation, although, recent reports have suggested activation of $TiO_2$ using visible light too [9].

Achieving antibacterial behaviour following above mentioned treatment however is often accompanied with associated risks [10]. In contrast, modifying surface properties of the implants appears to be a viable strategy while avoiding harmful effects of antibiotics. Hence, surface modification of implant material has emerged as a key area of research interest [11]. Among number of strategies to limit bacterial attachment on bio-material, surface texturing of implants using laser pulses is emerging as a promising option. This technique being a single step, chemical free process it is an attractive approach as it also allows surface treatment localized both, in time and space. A few recent studies have reported reduced bacterial attachment on superhydrophobic titanium surfaces limiting biofilm formation on the implants [12]. A number of fabrication techniques have been developed to mimic these superhydrophobic surfaces of which femtosecond laser based surface texturing is a recent one [13]. Laser based surface treatment provides a facile means of controlling surface properties of titanium alloys enabling fabrication of antibacterial surfaces. Various types of lasers have been used for surface treatment to achieve antibacterial surfaces. Although, surface structuring can be done using nanosecond and picoseconds lasers use of ultra-short femtosecond (fs) lasers is preferred as it offers several associated advantages. Femtosecond laser based surface modification is a highly precise and reproducible process, with minimal debris and heat affected zones, hence restricting collateral damage.

In this paper we present results of our investigation on antibacterial performance of Ti6Al4V alloy surface textured using fs-laser. At constant laser fluence levels Ti6Al4V alloy samples have been exposed to varying number of laser pulses by varying sample scanning speed in order to generate various laser induced surface topographies. A variety of surface micro-structures were generated varying from random to periodic self-aligned micro-conical structures post laser treatment. Results of two extreme cases of laser writing at slowest (25 µm/sec) and fastest (800 µm/sec) available scan speeds are discussed here. Detailed characterization of the laser modified surface included morphological study, change in surface chemical composition, wettability, surface roughness and investigation of surface crystalline phase. Results of all characterization tests are compared with those obtained for untreated Ti6Al4V alloy. Bacterial attachment tests were performed for three bacteria among which Staphylococcus aureus (S.

aureus) and Streptococcus mutans (S.mutans) were gram positive and Pseudomonas aeruginosa (P. aeruginosa) was gram negative bacteria. All these bacteria are clinically the most prevalent strains of bacteria responsible for hospital acquired and oral implant infections. The antibacterial behavior of Ti6Al4V alloy was tested and confirmed on Nutrient agar plate employing imprint method [14]. Scanning electron microscope (SEM) images showed attachment of bacteria on both untreated and laser treated surfaces. However, growth of bacteria was seen to occur only on untreated surface of Ti6Al4V alloy while a bacteria free zone was observed for laser treated sample. Absence of bacterial growth on the laser treated samples showed death of bacteria and hence confirmation of bactericidal property acquired post laser surface treatment of Ti6Al4V alloy. Our detailed investigations have correlated this bactericidal behavior to formation of oxide and sub-oxide phases of titanium supported by X-ray diffraction (XRD) and X-ray photoelectron spectroscopy (XPS) measurements. Enhanced surface roughness and formation of oxides on laser treatment appear to be responsible for the antibacterial behavior shown by the laser treated Ti6Al4V alloy surface.

Our study was aimed towards fabricating antibacterial and/or bactericidal titanium surfaces without use of antibiotics and UV-irradiation. Our results have successfully demonstrated laser treatment to be an effective approach for surface modification of titanium alloy based implants and biomedical samples for achieving superior antibacterial and bactericidal properties. Laser surface treatment therefore offers a potential alternative approach that does not involve use of antibiotics yet reduces implant associated infections.

## 2. Experiment

For our investigation, samples of Ti6Al4V alloy of dimension 10 mm X 10 mm X 1.2 mm were mechanically polished to surface roughness ~ 0.15 μm. The polished samples were subsequently cleaned in ultrasonic bath with acetone, ethanol and distilled water for 10 minutes each.

Cleaned Ti6Al4V sample surfaces were irradiated employing a Ti: Sapphire pulsed laser at a wavelength of 800 nm delivering pulses of 45 *fs* at a repetition rate of 3 kHz. Experimental setup used for laser surface texturing of Ti6Al4V alloy is illustrated in Fig. 1a. Experiments were performed in normal atmospheric environment (air) and humidity level of 60%. The laser beam was focused on the sample using a 50 cm focal length lens. Focal spot diameter of laser measured on the sample was ~ 300 µm. Typical average laser fluence of 0.6 J/cm$^2$ irradiated the sample surface. The samples were scanned at different scan speeds normal to the incident beam using a computer controlled X-Y stage. Surface morphology, chemical composition, and antibacterial behavior of two samples; Ti-1 and Ti-2 laser textured at scan speed of 25 µm/sec and 800 µm/sec, respectively are discussed here. Since the scan speed was slow for sample Ti-1, only 5 mm X 5mm area was laser treated on this sample. A photograph of sample Ti-1 laser treated at the centre (black region) surrounded with untreated region is shown in Fig. 1b. Unlike this, complete area of 10 mm X 10 mm was laser treated at faster scan speed of 800 µm/sec for sample Ti-2, its photograph is shown in Fig. 1c.

Surface topography of Ti6Al4V samples before and after laser treatment and after bacterial tests was observed under SEM (Carl Zeiss EVO 40 SEM). Grazing incidence XRD analysis of samples was carried out using PANalytical MRD system. The measurements were done using CuKα radiation of wavelength 1.54 A$^o$ in out of plane geometry over the 2θ range of $10^0$-$80^0$ at angle scanning speed of 2$^o$/min. XPS analysis of the Ti6Al4V samples before and after laser surface irradiation was performed using a VG make model CLAM-2 hemispherical analyzer with Al Kα source (1486.6eV) having a step size of 0.3 eV. Variation in wettability of Ti6Al4V surface after laser treatment was recorded by water contact angle measurement. The contact angle measurements were repeated at least 3 times for each sample. In vitro experiments were performed to study bacterial infection on laser treated Ti6Al4V alloy using nutrient agar plate as a broth. Details of the procedure of attachment and growth of bacteria on Nutrient broth/agar plate, incubation of samples and imprint process are similar to those reported in our earlier work [15].

## 3. Results :

### 3.1- Surface topography:

Figs. 2a and 2b are the SEM images of sample Ti-1 and sample Ti-2, respectively. Significantly different surface topography of the samples is distinctly visible in these micrographs. In Fig. 2a, wide laser tracks containing multi-domain structure are seen in the top view of sample Ti-1. A magnified image of sample Ti-1 shown in the inset of Fig. 2a shows porous surface of the protrusions. Unlike this, self aligned narrow conical structures were generated on sample Ti-2 surface, as shown in Fig. 2b and in its inset. Two major parameters which control laser-matter interaction are: laser power density and laser irradiation time. Such extensive difference in surface morphology observed in case of samples Ti-1 and Ti-2 was mainly due to a wide difference in cumulative laser irradiation time and Ti6Al4V surfaces. During the initial stage of interaction, micron level irregularities or roughness inherently present on the pristine Ti6Al4V alloy sample resulted in non-uniform deposition of laser energy on its surface. This resulted in generation of a temperature gradient on the sample surface. When the deposited energy on the sample was enough to initiate melting of the surface a liquid layer was formed on the surface. Surface tension driven flow of this liquid layer created a valley-peak structure. With each successive laser pulse irradiating the sample surface, these surface features grew, with peaks becoming prominent. Elongation of these protrusions largely depends on the amount of deposited energy and hence on the number of laser shots irradiating a particular spot. For sample Ti-1, the Ti6Al4V alloy was textured at slower scan speed of 25µm/sec that translated into ~ 36000 laser pulses/spot. However, sample Ti-2 scanned at much higher speed of 800µm/sec meant only ~1125 laser pulses/spot irradiated the sample. Therefore, the laser beam residence (interaction) time was longer for sample Ti-1 in comparison to sample Ti-2. The laser focal spot on the sample had a diameter 300 µm and lateral shift between two consecutive laser traces was 30 µm. Since, the lateral shift between two adjoining scan was less than the laser spot size, there was some amount of overlap between these laser tracks.

Figs. 3a, 3b and 3c show magnified (20 KX) SEM images of Ti-1, Ti-2 and untreated samples, respectively. In magnified scale, ridges of the protrusions of samples Ti-1 and Ti-2 are found to be covered with rippled structures. In contrast, the untreated sample had a very smooth surface. The summit density which is a count of number of peaks per unit area was estimated to be 3 and 10 for samples Ti-1 and Ti-2, respectively.

*3.2 – Chemical composition:*

Fig. 4a shows the XRD patterns of untreated Ti4V6Al sample. Distinct narrow peaks represent diffraction intensities at different angles from the untreated Ti4V6Al alloy. The measured peaks at angles - $35.5^o$, $38.7^o$, $40.6^o$, $53.3^o$, $63.8^o$ and $71.3^o$ coincide in position with Ti peaks. Phase determination of Ti6Al4V alloy suggested presence of hexagonal α-Ti (Joint Committee on Powder Diffraction Standards: JCPDS #44-1294) and cubic β-Ti phase (JCPDS #44-1288) [16]. In addition to titanium peaks, three more peaks appeared at diffraction angles- $20.9^o$, $23.1^o$, and $54.4^o$ for sample Ti-1, as shown in Fig. 4b. The peaks at $23.1^o$, and $54.4^o$ could be identified as $Ti_2O_3$ (JCPDS #89-4746) phase, while the most intense peak at $20.9^o$ matched well with the characteristic peak of $Ti_4O_7$ phase according to reference spectrum JCPDS #77-1391. Fig. 4c shows the XRD pattern of sample Ti-2. In this case peaks recorded at angles- $18.3^o$, $25.8^o$, $29.5^o$, $33^o$, and $56.6^o$ correspond to $Ti_3O_5$ (JCPDS #72-2101) phase, along with a single peak of Ti-O at $43.6^o$ (JCPDS #89-3660).

The surface chemistry analysis of Ti6Al4V alloy before and after laser treatment was carried out by XPS. Survey spectra of untreated Ti6Al4V alloy, sample Ti-1 and sample Ti-2 are shown in Fig. 5a, 5b and 5c, respectively. The survey spectra show presence of Carbon (C), Titanium (Ti) and Oxygen (O). Percentage of carbon decreased post-laser irradiation however, significant increase in oxygen content was observed on the laser treated surfaces. Increase in oxygen content could have occurred due to surface oxidation of samples during laser irradiation under atmospheric conditions.

Fig. 6a shows the high resolution XPS spectra of Titanium for: untreated, Ti-1 and Ti-2 samples. For untreated sample, the Ti2p spectra could be deconvoluted into two components having peak positions around 457.7eV and 462.0eV. These peaks correspond to $Ti_2O_3$ and 2p state of $Ti^{3+}$ valence state [17]. Spectra for sample Ti-1 was deconvoluted into three components with peak positions at 456.1eV, 459.0eV and 462.3eV. The peak appearing around 459.0eV is a characteristic peak of $Ti_4O_7$, while peaks at 456.1eV and 462.3eV can be associated with 2p3/2 and 2p1/2 valence states of $Ti^{3+}$. Spectrum of sample Ti-2 was deconvoluted into two peaks positioned at 458.5eV and 463.9eV, attributed to presence of $Ti^{4+}$ 2p3/2 and 2p1/2 valence states, respectively. Deconvolution of O1s spectra of untreated and Ti-2 samples resulted in two peaks, as shown in Fig. 6b. The peaks located at binding energy 530.05eV and 532.6eV for these samples correspond to $O^{-2}$ and C-O respectively. However, O1s spectra for sample Ti-1 when deconvoluted gave three components with peak positions at 530.2eV, 531.7eV and 534.0 eV, corresponding to $O^{-2}$, OH, and $H_2O$, respectively [18].

### *3.3- Surface wettability:*

Fig. 7a, 7b and 7c are the photographs of water contact angle measurement of the untreated, sample Ti-1 and sample Ti-2, respectively. The surface of untreated titanium alloy was hydrophilic with contact angle of 73±3°. Post laser treatment, wettability of the surface improved significantly with contact angle of 0° measured for sample Ti-1, as shown in Fig. 7b. In contrast, a time dependent reduction in the wettability was observed for sample Ti-2. For this sample, the surface was highly hydrophilic with water contact angle of 0° measured just after laser treatment. However, this wettability decreased and contact angle increased with time. Photographs of contact angle measurement of sample Ti-2 on different days are shown in Fig. 7d. The sample became hydrophobic on fourth day with contact angle of 99°, the surface becoming superhydrophobic after 10 days. Thereafter, the sample remained superhydrophobic contact angle remaining steady at a value of 160°. However, such variation in wettability for sample Ti-1 was not seen and it remained highly hydrophilic over a similar duration of time.

### *3.4- In-vitro testing:*

To study bacterial attachment and growth on laser treated Ti6Al4V alloy, in vitro experiments were performed on nutrient agar plate, used as a broth. Three different samples of type Ti-1 were immersed in three independent bacterial solutions: (a) S.aureus, (b) P.auroginosa, and (c) S.mutans and kept for 2 hrs each. Thereafter, imprint of each of these three samples were taken on nutrient agar plates which were subsequently incubated for 24 hours.

Figs. 8a, 8b, and 8c are SEM images of sample Ti-1, sample Ti-2 and untreated Ti6Al4V alloy sample dipped in S. aureus for 2 hrs. Attachment of S. aureus is clearly visible in all three images. One can notice that the superior hydrophilicity of sample Ti-1 surface supported more number of bacterial adhesion in comparison to other two.

Imprint images taken with nutrient agar plate showed extensive growth of all three bacteria, S.aureus, P.auruginosa, and S.mutans was found on untreated region of the sample Ti-1, as shown in Fig. 9a, 9b and 9c, respectively. A bacteria free zone was observed for S. aureus and S. mutans in Fig. 9a and 9c, respectively. However, few colonies of P. auroginosa were found on laser treated area, as shown in Fig. 8b.

In Figs. 10a-10c are shown photographs of nutrient agar plate indicating replica imprints of untreated Ti6Al4V alloy and sample Ti-2 immersed in S.aureus, P.auroginosa, and S.mutans bacterial solution, respectively. Untreated and laser treated samples are marked as UNT and T, respectively in the pictures. Sample Ti-2 surface allowed colonization of S.aureus bacterium. Adhesion and growth of this bacterium covering full surface of untreated and Ti-2 samples are evident in Fig. 10a. However, the sample Ti-2 rejected the growth of P. auruginosa and S. mutans completely as shown in Fig. 10b and 10c respectively. Growth of these bacteria on untreated sample is seen in Fig. 10b and 10c for comparison.

All our observations described above suggest that although all three types of bacteria attached on both untreated and laser treated surfaces, subsequent bacterial growth was strongly inhibited in all the three cases on the laser treated surface. Presence of oxides of Titanium post laser treatment could have played a critical role in inducing death of bacteria hence preventing bacterial growth as observed on nutrient agar plate images. Our observation showed no direct correlation between surface wettability of samples and their antibacterial behavior.

## 4. Discussion :

Varying interaction time between laser beam and Ti6Al4V alloy induced significantly different surface topography on two samples Ti-1 and Ti-2 as observed in present studies. Ji *et. al.* had reported strong attraction of S.aureus towards hydrophilic surface [19]. Fadeev e*t. al.* in their study found that S. aureus successfully colonized on superhydrophobic surfaces, while P. aureginosa cells could not attach to such surface [20]. These results are similar to our own observations. Presence of $Ti_4O_7$ megneli-phase of $TiO_2$ on Ti- sample has confirmed from our XRD and XPS analysis of Ti6Al4V samples post laser treatment. This $Ti_4O_7$ megneli phase has been reported to have high electric conductivity and great oxygen evolution potential [21]. In general, $Ti_4O_7$ phase of $TiO_2$ is obtained by annealing rutile phase of $TiO_2$ at high temperature ($< 1000^o$ C) in presence of $H_2$. We observed generation of $Ti_4O_7$ phase on fs-laser treated sample Ti-1. It is reported that, $Ti_4O_7$ bonds to OH loosely in water thereby assisting in dimerization of OH leading to generation of hydrogen peroxide ($H_2O_2$) [22]. $H_2O_2$ has been known for its extensive use as a biocide since reactive oxygen species associated with $H_2O_2$ kills bacteria. Hence, presence of $Ti_4O_7$ on the surface of sample Ti-1 could have successfully inhibited growth of all three bacteria as observed in our study.

As observed in Fig. 10a-10c, while, sample Ti-2 completely inhibited growth of P. auruginosa and S. mutans it could not stop colonization and growth of S. aureus. This could have occurred on account of two probable reasons. S. aureus bacterium is spherical in shape with size around 0.5-2.0 µm. P. auruginosa is rod shape with size 1.0-5.0 µm and S.mutans are long chain structure with size ranging from 1.0 - 4.0 µm. Higher summit density and increased surface roughness for sample Ti-2 could have resulted in discontinuous contacts of surface features with longer bacteria. This reduced the bonding opportunity of these two bacteria and restricted their contact with the titanium alloy surface. However, S. aureus with its small spherical ball like structure could penetrate into the valley and settle there enabling them to colonize. Also, $Ti_3O_5$ and TiO phases of titanium oxide were detected on sample Ti-2 post laser treatment unlike sample Ti-1 where more reactive $Ti_4O_7$ had been formed. While, growth of P.auruginosa and S. mutans could be inhibited by these less reactive oxides of titanium S. aureus managed to survive in this environment of $Ti_3O_5$ and TiO. This observation is also supported by the fact that S. aureus has been reported to survive by developing defenses against oxidative stress [23]. This makes S. aureus less susceptible to toxicity associated with $Ti_3O_5$ and $TiO_2$ in comparison to the other two bacteria.

## 5. Conclusion :

We have achieved two distinct surface morphologies on Ti6Al4V alloy using femtosecond direct laser writing technique by changing sample scan speed. Along with surface morphology, surface chemical compositions of such samples were also found to be different. Our XPS results demonstrated the formation of $Ti^{+3}$ and $Ti^{+4}$ valence states on sample Ti-1 and sample Ti-2, respectively, post laser treatment. Formation of $Ti_4O_7$ and $Ti_3O_5$ major phases on sample TI-1 and sample Ti-2, respectively, were also confirmed by XRD characterization of samples. Samples laser treated under optimized conditions showed superior antibacterial behavior against S. aureus, P. aeruginosa and S. mutans in comparison to untreated Ti6Al4V alloy. Thus, surface treatment of Ti6Al4V alloy using fs-laser appears to be an efficient technique for obtaining antibacterial and/or bactericidal Ti6Al4V alloy surfaces without necessitating use of any antibiotics or UV-illumination.

**Acknowledgment:** We acknowledge Ms. Sridevi and Mrs.Nidhi gupta, Technical Physics Division, Bhabha Atomic Research Centre, Mumbai 400085 India, for technical help.

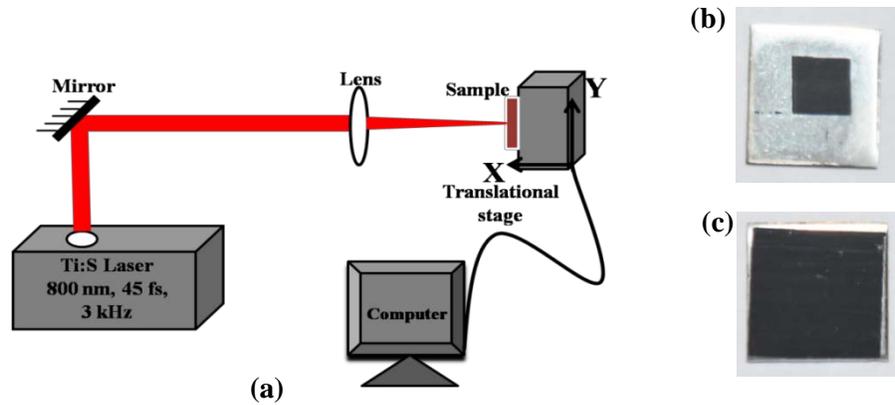

Figure 1 (a) Experimental setup for fs-laser treatment of Ti6Al4V samples, photograph of (b) sample Ti-1 and (c) sample Ti-2

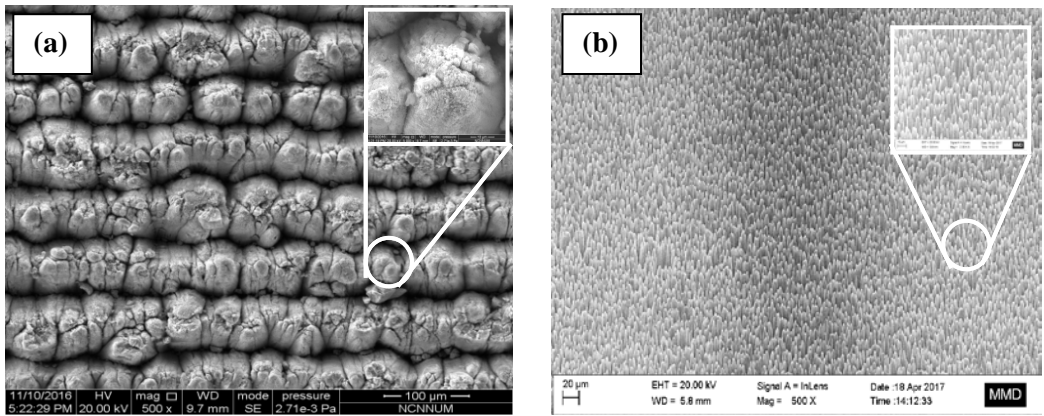

Figure-2: SEM image of (a) sample Ti-1, and (b) sample Ti-2, insets: corresponding top view at magnification of 2KX

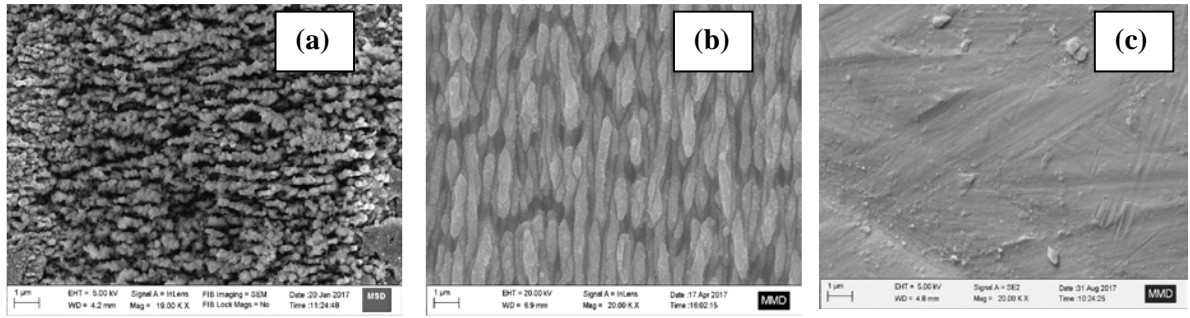

Figure-3: SEM image of (a) sample Ti-1, (b) sample Ti-2 and (c) untreated sample, at magnification of 20 KX

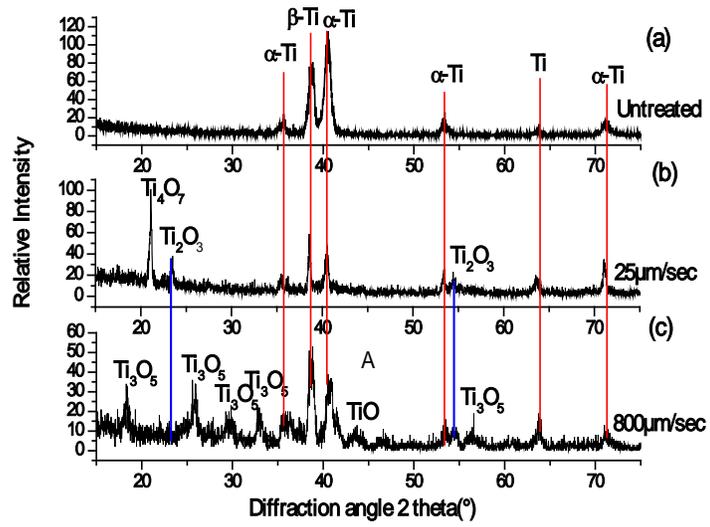

Figure-4: XRD pattern of (a) untreated, (b) Ti-1 and (c) Ti-2 samples

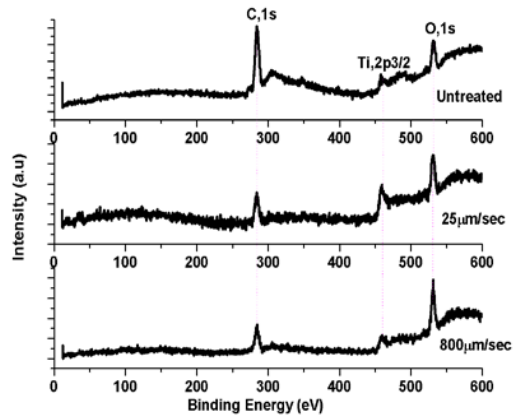

Figure-5: XPS survey spectra of (a) untreated, (b) Ti-1 and (c) Ti-2, samples

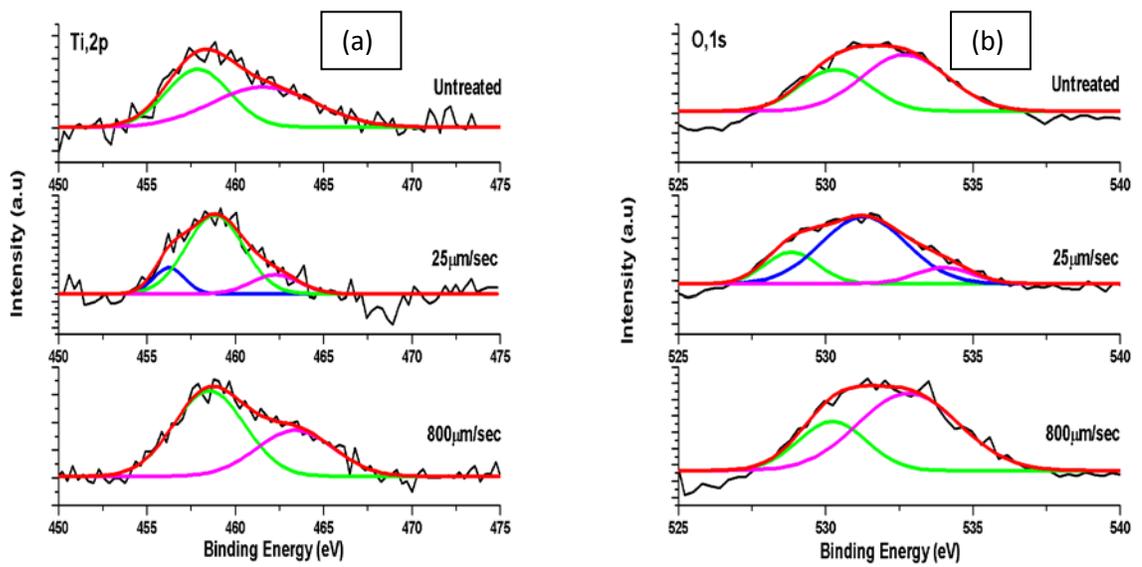

Figure-6: High resolution XPS spectra for Ti6Al4V alloy samples, Ti-1 and Ti-2 samples (a) Ti2P and (b) O1s

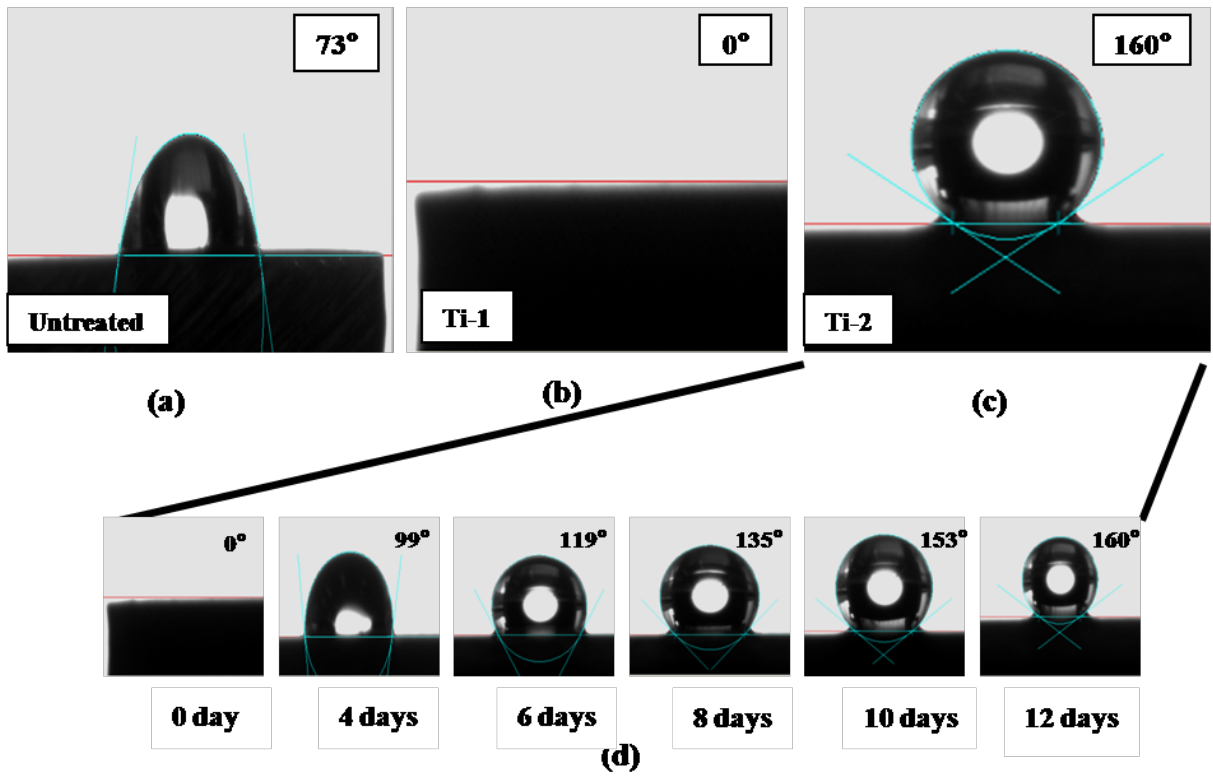

Figure-7: Photographs of water contact angle measured on (a) untreated, laser treated (b) Ti-1, (c) Ti-2 and (d) time dependent variation in water contact angle of laser treated implant surface of Ti-2.

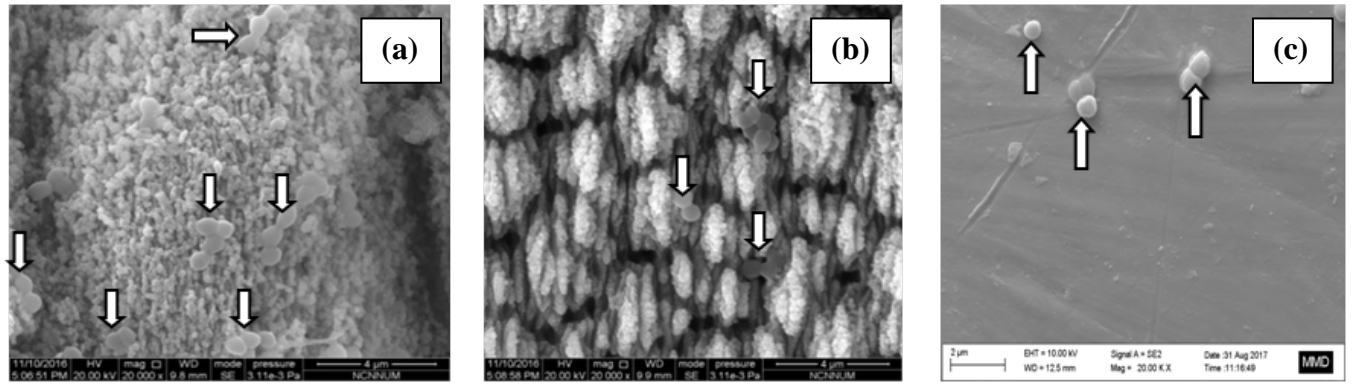

Figure-8: FE-SEM images of S.aureus cultured on (a) sample Ti-1, (b) sample Ti-2 and (c) untreated Ti6Al4V alloy surfaces

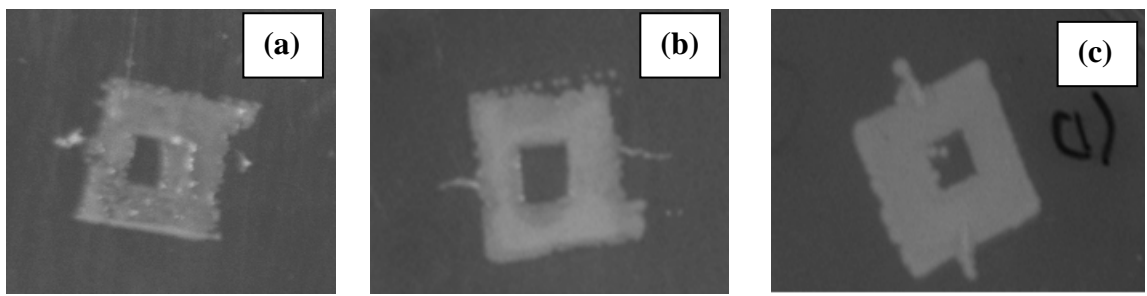

Figure-9: Photographs of nutrient agar plate showing replica imprint of titanium implant with 5 mm X 5mm square zone laser treated at scan speed 25μm/sec (Ti-1) immersed in (a) S.aureus ,(b) P.auruginosa , and (c) S.mutans bacterial solution for 2 hours

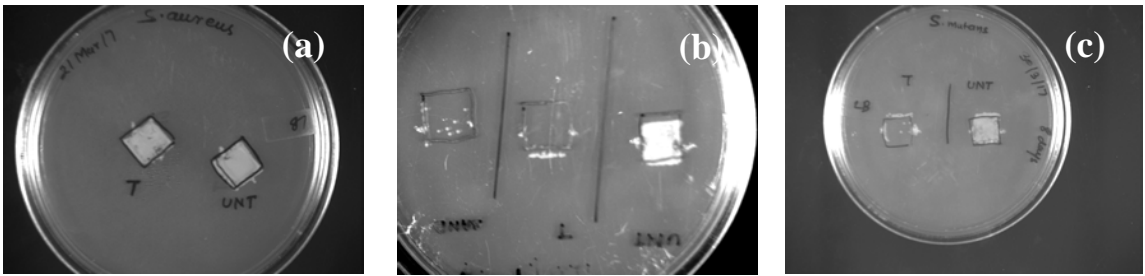

Figure-10: Photographs of nutrient agar plate showing replica imprint of Ti-2 sample immersed in (a) S.aureus, (b) P.auruginosa, and (c) S.mutans bacterial solution for 2 hours